\documentclass[aps,pra,twocolumn,showpacs,floatfix]{revtex4}

\begin{document}

\title{Non-empirical hyper-generalized-gradient functionals constructed from 
the Lieb-Oxford bound}

\author{Mariana M. Odashima and K. Capelle}
\affiliation{Departamento de F\'{\i}sica e Inform\'{a}tica,
Instituto de F\'{\i}sica de S\~ao Carlos, Universidade de S\~ao Paulo,
Caixa Postal 369, 13560-970 S\~ao Carlos, SP, Brazil}
\date{\today}

\begin{abstract}
A simple and completely general representation of the exact 
exchange-correlation functional of density-functional theory is derived from
the universal Lieb-Oxford bound, which holds for any Coulomb-interacting 
system. This representation leads to an alternative point of view on popular 
hybrid functionals, providing a rationale for why they work and how they can 
be constructed. A similar representation of the exact correlation functional
allows to construct fully non-empirical hyper-generalized-gradient 
approximations (HGGAs), radically departing from established paradigms of 
functional construction. Numerical tests of these HGGAs for atomic and 
molecular correlation energies and molecular atomization energies show that
even simple HGGAs match or outperform state-of-the-art correlation functionals 
currently used in solid-state physics and quantum chemistry.
\end{abstract}

% 31.15.eg Exchange-correlation functionals (in current density functional theory)
% 31.15.V- Electron correlation calculations for atoms, ions and molecules
% 71.15.Mb Density functional theory, local density approximation
% 71.10.Ca Electron gas, Fermi gas

\maketitle

\newcommand{\be}{\begin{equation}}
\newcommand{\ee}{\end{equation}}
\newcommand{\bea}{\begin{eqnarray}}
\newcommand{\eea}{\end{eqnarray}}
\newcommand{\bi}{\bibitem}
\newcommand{\la}{\langle}
\newcommand{\ra}{\rangle}
\newcommand{\ua}{\uparrow}
\newcommand{\da}{\downarrow}
\renewcommand{\r}{({\bf r})}
\newcommand{\rp}{({\bf r'})}
\newcommand{\rpp}{({\bf r''})}

\section{Introduction}
\label{intro}

The modern understanding of materials is based on the detailed quantitative 
description of their electronic structure afforded by quantum-mechanics.
Since electron-electron interactions play an essential role in shaping 
material properties, electronic structure must be dealt simultaneously with 
the many-body physics of interacting electrons. At the heart of this interface 
between many-body theory and electronic-structure theory is the 
exchange-correlation ($xc$) functional of density-functional theory (DFT) 
\cite{kohnrmp,dftbook,parryang,perdewoverview}. 

This functional, $E_{xc}[n]$, comprises the exchange energy $E_x[n]$, which
is known in terms of single-particle orbitals, and the correlation energy 
$E_c[n]$, which is unknown, and must be approximated.
An intense and multidisciplinary research effort is thus directed at
developing ever better approximations for $E_{xc}[n]$ or $E_c[n]$.
While the local-density approximation (LDA) \cite{pw92} is still widely and 
successfully used for band-structure calculations, total-energy calculations 
require at least the accuracy afforded by generalized-gradient approximations 
(GGAs) \cite{pbe,pw91,lyp}. 
Quantum-chemical applications of DFT to atoms and molecules make 
increasingly use of hybrid functionals \cite{hyb0,hyb2,hyb3} mixing a certain
(often empirical) fraction of exact exchange with LDA or GGA. 

To make progress beyond the GGA level, the concepts of meta-GGA (MGGA) 
\cite{pkzb,tpss} and hyper-GGA (HGGA) \cite{PSTS,MCY,beckehgga} 
have been proposed for 
functionals employing the kinetic-energy density or the exchange-energy 
density, respectively, but few explict examples of such 
functionals have been constructed. Established procedures of functional 
construction do not fully determine the form of GGAs, MGGAs and HGGAs, 
enhancing the need for introducing empirical parameters. 

Here we propose a hitherto unexplored mode of functional
construction, which provides additional insight into the structure of the 
exact $xc$ functional and the nature of common approximations to it, and 
naturally gives rise to non-empirical hyper-GGA functionals.

In Sec.~\ref{representations} we use the universal Lieb-Oxford (LO) bound
\cite{lo} on 
the exchange-correlation energy of Coulomb-interacting systems to derive an
exact representation of the universal exchange-correlation functional of
DFT. As an immediate consequence, we also obtain an LO-based exact 
representation of the universal correlation-energy functional.

In Sec.~\ref{hybrids}, the LO-based representation of $E_{xc}[n]$ is shown 
to lead naturally to a reconstruction of the generic global hybrid functional,
providing additional insight into the meaning of the components of the hybrid.
Local hybrids are shown to be describable in a similar way.

Section~\ref{hggaconstruction} is devoted to using the LO-based representation 
of $E_c[n]$ to construct a family of orbital-dependent correlation functionals
that in the Jacob's ladder classification scheme \cite{perdewoverview} belong
on the hyper-generalized-gradient approximtion (HGGA) rung. The resulting
HGGA functional is of different form from other functionals on the same rung,
and satisfies exact constraints such as scaling properties and recovery of
the gradient expansion for weakly varying densities.

In Sec.~\ref{hgga1tests} we report numerical tests of the constructed HGGA for
atomic and molecular correlation energies and molecular atomization energies,
and compare to common correlation functionals of the LDA, GGA and meta-GGA 
type. We also present, in Sec.~\ref{variants}, two further LO-based HGGA 
correlation functionals, which perform better than that of the preceding 
section for specific 
properties. One of these employs an unusual multiplicative self-interaction
correction, the other enhances the possibilities for error cancellation in
combination with semilocal approximations for exchange.

Section~\ref{concl} contains our conclusions.

\section{Lieb-Oxford-based representations of the correlation and 
exchange-correlation energies}
\label{representations}

The starting point of our analysis of $E_c$ is the Lieb-Oxford (LO) bound
\cite{lo}, according to which the exchange-correlation energy obeys
\be
E_{xc}[n] \ge \lambda E_x^{LDA}[n],
\label{lo2}
\ee
where the Lieb-Oxford value for $\lambda$ is $\lambda_{LO}=2.27$. 
Recent numerical evidence suggested that the LO 
bound can be further tightened \cite{ch,lotighten} and generalized to
one and two-dimensional systems \cite{lowdim}. We therefore 
also report results obtained from the value $\lambda_{EL}=1.9555 
\approx 2.0$, which is the ratio of $E_{xc}$ to $E_x^{LDA}$ in the extreme 
low-density limit of the uniform electron liquid (EL), which was
conjectured to provide the tightest universally applicable bound 
\cite{lotighten,lowdim}.

In the form (\ref{lo2}), the LO bound plays a key role in the development of 
some GGAs \cite{pbe,revpbe,rpbe} and meta-GGAs \cite{pkzb,tpss}. 
The form of these functionals is dictated by other 
considerations, but typically the value of an otherwise undetermined parameter 
in them is chosen such that the bound is obeyed for all possible densities. 
Here we make rather different use of the LO bound: instead of using it to
fix the value of a parameter in a functional whose form is obtained in 
other ways, we use it to determine the form of the functional itself.

By combining the variational principle with the LO bound in its form 
(\ref{lo2}), we find immediately
\be
0 \ge E_x[n] \ge E_{xc}[n]  \ge \lambda E_x^{LDA}[n].
\ee
$E_{xc}$ is thus bounded from above and below, which allows us to cast it as
\be
E_{xc}[n] = (1-\beta[n]) E_x[n] + \beta[n]\lambda E_x^{LDA}[n],
\label{rep1}
\ee
where $\beta[n]$ is a density functional taking values in the interval 
$[0,1]$. By subtracting $E_x[n]$ we find for the correlation functional 
the exact representation
\be
E_c[n] = \beta[n] \left(\lambda E_x^{LDA}[n] - E_x[n]\right),
\label{rep2}
\ee
Since these representations of $E_{xc}$ and $E_c$ are completely 
general, all exact constraints on these functionals become constraints 
on $\beta[n]$, with the 
difference that while $E_{xc}[n]$ and $E_c[n]$ have values ranging, in 
principle, from zero to $-\infty$, $\beta[n]$ ranges from zero to one.

Although in principle $\beta[n]$ is as complex as $E_{xc}[n]$, we note that 
the maximum absolute error one can make in approximating a quantity varying
from $0$ to $-\infty$ is $\infty$, while for a quantity varying from $0$ to
$1$ is is $1/2$. This observation suggests that it may be useful to develop
simple models for $\beta[n]$, designed to recover as many exact properties 
as possible, in order to develop new approximations for $E_{xc}[n]$.

\section{Connection to global and local hybrids}
\label{hybrids}

Representations (\ref{rep1}) and (\ref{rep2}) have many interesting properties,
of which we now explore a few. As a first application, we compare 
Eq.~(\ref{rep1}) with hybrid functionals, of which a typical single-parameter 
example can be written as \cite{hyb0,hyb2,hyb3}
\be
E_{xc}^{hyb}[n] = (1-a) E_x[n] + a E_x^{LDA}[n] + E_c^{approx}[n].
\label{hyb}
\ee
Here $a$ is a (normally empirical) constant determining how much LDA 
exchange is mixed into exact exchange.

Expression (\ref{hyb}) can be considered an approximation to the exact 
representation (\ref{rep1}), consisting of three steps: (i) Replace 
the functional $\beta[n] \in [0,1]$ by a parameter $a \in [0,1]$. 
(ii) Replace the LO parameter $\lambda$ by unity, in the second term. 
According to the general LO bound, this means that the correlation energy 
is underestimated, {\em i.e.}, the resulting energy is not low 
enough. (iii) This underestimate is compensated by adding an explicit 
correlation functional $E_c^{approx} \leq 0$.

This re-construction of the generic hybrid (\ref{hyb}) starting from the
exact representation (\ref{rep1}) suggests an alternative interpretation
of the individual contributions to the hybrid functional: $E_c^{approx}$ 
is not an approximation to the full correlation energy, but only 
to the part missed by replacing $\lambda=2.27$ by $1$. We stress that this is 
merely a change in perspective, as the final form is exactly the same. 
However, such a change may be useful in selecting correlation functionals 
to be used in conjunction with exchange hybrids, and in the construction of 
novel hybrids. 

A related class of functionals, so-called local hybrids, are of the generic 
form \cite{lhyb}
\bea
E_{xc}^{lhyb}[n] = \int d^3 r [(1-a\r) e_x[n]\r + a\r e_x^{LDA}[n]\r 
\nonumber \\ 
+ e_c^{approx}[n]\r],
\label{lhyb}
\eea
where lower-case letters indicate energy densities.
These functionals, too, can be interpreted as particular approximations to
the general LO-based representations, by starting from the {\em local} 
LO bound, $ e_{xc}[n]\r \ge \lambda e_x^{LDA}[n]\r$. This local form of 
the bound (satisfaction of which guarantees satisfaction of the global one)
is that also used in constructing PBE-GGA and TPSS meta-GGA. In terms of
this bound, the different components of the local hybrid can be interpreted 
in the same way as for the global hybrid. However, we note that unlike the 
global LO bound, the local one is not a rigorous property of quantum mechanics, 
but may be violated. Thus, in this sense, local hybrids are less tightly 
connected to universal bounds than global hybrids.

\section{Construction of a hyper GGA}
\label{hggaconstruction}

As a second application, we use representation (\ref{rep2}) to construct a
class of non-empirical hyper-GGAs by enforcing constraints on $\beta[n]$.
Occasionally, the expression hyper-GGA is meant to refer to any functional
employing exact exchange. In this sense, the global and local hybrids just
discussed are already hyper-GGAs. However, we here adopt a more restrictive
definition, in which hyper-GGA refers specifically to correlation functionals
that use the exchange-energy density as an ingredient.

A first approximation to the functional $\beta[n]$ is obtained by requiring
that the resulting $E_c[n]$ has the correct uniform density limit.
On uniform densities $\bar{n}$ the exact and general
representation (\ref{rep2}) becomes
\be
E_c[\bar{n}] =
\beta[\bar{n}] \left(\lambda E_x^{LDA}[\bar{n}] - E_x[\bar{n}]\right),
\ee
and since by definition $E_c[\bar{n}] = E^{LDA}_c[\bar{n}]$ and
$E_x[\bar{n}] = E^{LDA}_x[\bar{n}]$, we find
\be
\beta[\bar{n}]=\frac{E^{LDA}_c[\bar{n}]}{(\lambda-1)E^{LDA}_x[\bar{n}]}.
\ee
The use of this $\beta$ in Eq.~(\ref{rep2}) also for nonuniform densities
leads to a LO-based functional that correctly recovers the uniform gas limit.

\begin{table*}
\begin{ruledtabular}
\caption{\label{table1} Comparison of our HGGA1 functional with standard LDA,
GGA and MGGA correlation functionals. First row: negative correlation energy
of the H atom, in milliHartree.  Second row: mean absolute relative error
(mare) of the correlation energy of 17 atoms, from He ($Z=2$) to Ar ($Z=18$).
Third row: mare of the correlation energy of 35 molecules for which highly
precise correlation energies are available \cite{molrefs}.
Fourth row: correct for the electron liquid (y/n).}
\begin{tabular}{lccccccccc}
&LDA & PW91 & PBE & LYP & TPSS &
HGGA1($\lambda_{LO}$)&
HGGA1($\lambda_{LO}$)&
HGGA1($\lambda_{EL}$)&
HGGA1($\lambda_{EL}$)\\
&&&&&&&MSIC &&MSIC \\
\hline
H atom (mH) & 21.66 & 6.33 & 5.71 & [0] & [0] & 6.24 & 0 & 6.20 & 0 \\
atoms (mare \%)& 119.7 & 4.9 & 6.8 & 3.9 & 5.4 & 4.75 & 4.38 & 4.70 & 4.37 \\
molecules (mare \%)& 102.5 & 7.4 & 9.7 & 6.7 & 9.0 & 7.0 & 6.6 & 6.8 & 6.5 \\
electron liquid& Y & Y &Y &N &Y &Y &Y &Y &Y \\
\end{tabular}
\end{ruledtabular}
\end{table*}

In the same way, we can build in the gradient expansion for weakly
nonuniform densities $\tilde{n}\r$, where the tilde means that the density
is such that the low-order gradient-expansion approximation (GEA) is 
adequate. On such densities representation (\ref{rep2}) becomes
\be
E_c[\tilde{n}] =
\beta[\tilde{n}] \left(\lambda E_x^{LDA}[\tilde{n}] - E_x[\tilde{n}]\right),
\ee
and since by definition $E_c[\tilde{n}] = E^{GEA}_c[\tilde{n}]$ and
$E_x[\tilde{n}] = E^{GEA}_x[\tilde{n}]$, we can construct an approximation 
to $\beta[n]$ from any functional that on weakly varying densities correctly 
reduces to the gradient expansion. The simplest choice would be the gradient 
expansion itself, leading to 
\be
\beta[\tilde{n}]=
\frac{E^{GEA}_c[\tilde{n}]}{\lambda E^{LDA}_x[\tilde{n}]-E_x^{GEA}[\tilde{n}]}.
\ee
The use of this $\beta$ in Eq.~(\ref{rep2}) also for strongly nonuniform 
densities leads to an explicit functional correlation functional 
recovering the uniform and the weakly nonuniform limits, in addition to 
the LO bound.

However, the denominator of this prefactor can have zeros at some values
of the reduced gradient $s$. The prefactor $\beta[n]$ diverges at such
densities, in contradiction to the Lieb-Oxford bound. Thus, this bound 
requires to construct $\beta$ from a functional that has the correct gradient 
expansion at small gradients, and does not produce divergences at large ones. 
This problem is solved by using GGA instead of GEA, {\em i.e.},
\be
\beta[\tilde{n}]=
\frac{E_c^{GGA}[\tilde{n}]}{\lambda E_x^{LDA}[\tilde{n}]-E_x^{GGA}[\tilde{n}]},
\label{betagga}
\ee
which is also exact for weakly varying densities.
The use of this $\beta$ in Eq.~(\ref{rep2}) for arbitrary
densities leads to the approximate functional
\be
E_c^{HGGA1}[n]=
\frac{E_c^{GGA}[n]}{\lambda E_x^{LDA}[n]-E_x^{GGA}[n]} 
\left(\lambda E_x^{LDA}[n]-E_x[n]\right).
\label{hgga1}
\ee

By construction, this functional recovers the uniform limit and the 
gradient expansion {\em to the order built into GGA}. This is an important
caveat, since actually very few GGAs recover the GEA both for exchange 
and correlation. In fact, the requirement that both the exchange and 
correlation functional used in constructing $\beta[n]$ reproduce the 
low-order gradient expansion excludes popular GGAs such as PBE, PBEsol and 
BLYP, and almost uniquely singles out PW91 as the only widely used GGA 
suitable for the construction \cite{footnote1}. 

Numerically, we have explored many other variations, employing different 
choices of ingredient functionals of $\beta[n]$ that do not fully recover 
the gradient expansion, such as PBE. This empirical analysis confirms that 
the choice of PW91 in $\beta[n]$ is near-optimal (and certainly better than 
PBE) independently of, but in agreement with, the above construction based 
on recovering exact constraints.  

With this choice, Eq.~(\ref{hgga1}) has become an explicit correlation
functional expressed in terms of other known density functionals, in 
particular the exact exchange functional $E_x$. For this reason, it belongs 
into the class of hyper-GGAs. Interestingly, while representations 
(\ref{rep1}) and (\ref{rep2}) thus rather naturally lead to a connection 
with hybrids and to hyper-GGA type functionals, they do not involve any 
explicit use of kinetic energy densities, {\em i.e.}, the present functionals 
belong to the fourth rung of Jacob's ladder \cite{perdewoverview,PSTS,kuemmel} 
without having passed through the third (meta-GGA) rung.

We note that the LO bound is incorporated in the correlation functional 
(\ref{hgga1}) through its structure, not by choice of a parameter, as in 
common GGAs and MGGAs. It also makes use of the bound for correlation, unlike
PW91 GGA, PBE GGA and TPSS meta-GGA, which use it for exchange. Moreover, 
Eq.~(\ref{hgga1}) employs only the {\em global} LO bound, which has been 
rigorously proven for all possible densities \cite{lo}, and does not make 
use of the {\em local} LO bound.

We also observe that under Levy coordinate scaling \cite{levyscaling}
$n\r \to n_\gamma\r = \gamma^3 n(\gamma {\bf r})$, both $E_x^{LDA}$ and 
$E_x^{PW91}$ correctly scale as $E_x[n_\gamma]=\gamma E_x[n]$. Thus, on 
scaling the present $E_c^{HGGA1}[n]$ the exchange scaling factors cancel,
and $E_c^{HGGA1}[n]$ appropriately inherits all scaling properties of 
the correlation functional used in $\beta$. 

Overall, Eq.~(\ref{hgga1}) is a nonempirical (in the 
sense of not containing any fitted parameters) correlation functional that 
recovers the uniform and the weakly nonuniform limits by construction, 
correctly scales as a correlation functional, and obeys the LO bound for 
structural reasons. 

On the down side, in its present form the functional is not size
consistent. However, we note that size-consistent versions of all our
functionals can be constructed by replacing the component energy functionals
by their respective energy densities, and integrating over the result,
very similar to the step leading from a global hybrid to a local hybrid.
In the specific case of HGGA1, this leads to
\bea
E_c^{HGGA1-SC}[n]= \nonumber \\
\int d^3r\, \frac{e_c^{GGA}[n]\r \left(\lambda e_x^{LDA}[n]\r-e_x[n]\r\right)}
{\lambda e_x^{LDA}[n]\r-e_x^{GGA}[n]\r},
\eea
which is size consistent (but, as any other approximate density functional 
available, not guaranteed to be size extensive).

\section{Numerical tests}
\label{tests}

\begin{table*}
\begin{ruledtabular}
\caption{\label{table2} Atomization energies of the set of 20
molecules used in Ref.~\cite{pbe}. Values are mean average errors
in kcal/mol, and were obtained from spin-unrestricted calculations.
First row: Using $E_c^{\rm approx}$ combined with exact (Fock) exchange.
Second row: Using $E_c^{\rm approx}$ combining each approximation with the
corresponding conventional choice for approximate exchange and all HGGAs
with $E_x^{PW91}$.}
\begin{tabular}{lcccccccc}
&LDA & PW91 & PBE & BLYP & TPSS & HGGA1& HGGA1$\cdot$MSIC& HGGA2\\
\hline
with $E_x$ & 30.8 & 29.4& 29.3 & 29.1& 26.8& 28.3& 23.7& 33.3 \\
with $E_x^{\rm approx}$ & 31.5 & 8.0 &  7.8 & 4.5 & 3.1 & 9.1 & 13.7& 4.0
\end{tabular}
\end{ruledtabular}
\end{table*}

\subsection{HGGA1 correlation functional}
\label{hgga1tests}

In this section we present numerical tests of our HGGA1 functional. Since
this is a correlation functional, the most direct and stringent test is 
the calculation of atomic and molecular correlation energies, to which we
turn first.

A selfconsistent implementation of orbital functionals such as 
Eq.~(\ref{hgga1}) is possible by means of the optimized effective 
potential method and its simplifications \cite{kuemmel}, or by 
the scaled selfconsistency approach \cite{ssc}. 
Alternatively, such functionals can be implemented post-selfconsistently on 
LDA or GGA densities. Below, we evaluate all component-functionals of our
HGGAs post-selfconsistently on PW91 densities and orbitals.

Table \ref{table1} compares our HGGA1 correlation functional 
to a set of widely used correlation functionals: LDA, PW91 GGA, PBE GGA, 
LYP GGA, and TPSS MGGA, for 18 atoms, 35 molecules and the electron liquid. 
All DFT calculations were performed with {\tt GAUSSIAN 03} 
\cite{gaussian} using the aug-cc-pVQZ basis set (except for the molecules from
Ref. \cite{MantenLuchow}, which uses cc-pVTZ).
 
The first row of Table \ref{table1} reports correlation energies of the 
hydrogen atom, which we display separately from those of other atoms because 
they are exclusively due to self-interaction, and thus permit to assess the
self-interaction error. We note that already HGGA1, without any
self-interaction correction, has a lower self-correlation error than LDA and 
PW91 GGA. HGGA1$\cdot$MSIC, which by construction has zero self-correlation 
error, is explained in Sec.~\ref{variants}, below.

The second row reports mean absolute relative errors (mare) for atoms He to 
Ar. 
As benchmark data we used the standard set of CI atomic correlation energies 
of Ref. \cite{froese}. HGGA1 performes better than LDA, PW91, PBE and even 
TPSS MGGA, but loses to LYP. LYP, unlike all the other functionals tested
here, contains empirical parameters fitted to the He atom, which explains its 
superior performance when applied to isolated atoms. 

The third row reports molecular correlation energies for a set of 35 molecules
for which highly precise correlation energies are available \cite{molrefs}.
Encouragingly, we find that on this set HGGA1 achieves a lower error than all 
tested nonempirical functionals, including the highly sophisticated TPSS MGGA.
The fourth row reports whether the tested functional is correct for the
uniform electron liquid. All correlation functionals except for LYP (which
was not designed to be correct in this limit) pass this test.
Finally, we note that the performance of HGGA1 is systematically improved 
by lowering the value of $\lambda$. This improvement is
particularly encouraging, as a constraint-based functional should indeed
deliver better results when the constraint it is based on is sharpened.

While our functional is, by construction, a correlation functional, and 
thus should be, as a matter of principle, compared to other correlation
functionals, in practice it is clearly important to also test its performance 
for molecular atomization energies. As a test set we employed the well 
established set of 20 molecules that was used in the original PBE work 
\cite{pbe}.
Atomization energies are calculated from total energies, which in turn require 
chosing an exchange functional in addition to a correlation functional.
To compare like with like, we combine all correlation functionals 
included in Table~\ref{table1} with the same exchange functional. Since
the spirit of HGGA\cite{kuemmel} is to provide a correlation
functional to be combined with exact (Fock) exchange, we first calculated all
total and atomization energies combining $E_c^{\rm approx}$ with the exact
$E_x$. Results are reported in the first line of Table~\ref{table2}.
HGGA1 performs better than LDA, PW91, PBE and BLYP, while HGGA1$\cdot$MSIC 
also improves on TPSS. When combined with exact exchange, the present HGGA
functionals thus provide, comparatively, the best atomization energies.

\subsection{Variants exploiting error cancelation and the one-electron limit}
\label{variants}

On the other hand, it is well known that error cancellation allows to obtain
much better atomization energies from local and semi-local correlation 
functionals by combining them with approximate local or semilocal exchange 
functionals, instead of with exact exchange. Results obtained in this way are 
reported in the second line of Table~\ref{table2}, where LDA, PW91, PBE and 
TPSS correlation were combined with the corresponding exchange functionals,
LYP correlation was combined with B88 exchange, and all HGGA's were
combined with PW91 exchange. 

Evidently, all semilocal functionals benefit hugely from this error 
cancellation. The same applies to HGGA1, but the degree of improvement
is slightly smaller than for GGAs and MGGA. However, we note that our choice of
$\beta[n]$ in the construction of HGGA1 is not unique. The particular
combination leading to HGGA1 was determined by the requirement to recover
the gradient expansion, which turns out to produce rather good 
{\em stand-alone} correlation energies. 

In order to provide a fair comparison with 
functionals exploiting error cancellation, we constructed an alternative 
HGGA that has larger correlation-energy errors than HGGA1, and does not
recover the gradient expansion, but combines better with semilocal exchange. 
This functional,
\be
E_c^{HGGA2}[n]=
\frac{E_c^{LDA}[n](\lambda E_x^{LDA}[n] - E_x[n])}
{\lambda E_x^{PW91}[n]-E_x[n]}.
\label{hgga2}
\ee
is denoted HGGA2 in Table~\ref{table2}, and found to give better atomization 
energies than all other tested functionals, except for the very sophisticated 
TPSS MGGA. It should not, however, be used to calculate correlation energies,
for which it performs much worse than HGGA1. This behaviour is similar to
that of common hybrids, which give excellent atomization energies but much
worse correlation energies.

While HGGA2 is a variation designed to benefit from error cancellation with
a semilocal exchange functional, we can also build in other desirable features
directly on top of HGGA1. As an example, we build in
a novel type of self-interaction correction (SIC).
Just as the property $E_x[\bar{n}]=E_x^{LDA}[\bar{n}]$ serves as an integrated
homogeneity indicator, the property $E_x[n^{(1)}]=-E_H[n^{(1)}]$, where
$n^{(1)}$ is any one-electron density, is an integrated one-electron indicator.
We build this indicator into HGGA1 by multiplying Eq.~(\ref{hgga1}) with the 
multiplicative SIC (MSIC) factor \cite{footnote2}
\bea
F_{MSIC}= \frac{E_x[n]+E_H[n]}{E_x^{LDA}[n]+E_H[n]},
\label{msic}
\eea
which was designed to yield zero for one-electron densities. 
LYP GGA and TPSS MGGA also achieve this, but at a price: LYP erroneously
predicts zero correlation energy for any fully polarized system, while
the local one-electron indicator used in TPSS MGGA
\cite{tpss} recognizes only one-electron systems with real orbitals,
but fails for complex (current-carrying) orbitals. Thus, the correct entry 0
for TPSS MGGA, and even more so that for LYP GGA, must be interpreted with
caution, signaled in Table \ref{table1} by $[0]$. The global one-electron 
indicator $E_x[n^{(1)}]=-E_H[n^{(1)}]$, used in our MSIC, does not suffer 
from either problem.

The resulting
HGGA1$\cdot$MSIC functional is a product of three factors, one steming
directly from the LO bound, one ($\beta$) from the electron liquid, and
one ($F$) from the one-electron limit. HGGA1$\cdot$MSIC spoils the 
recovery of the gradient expansion achieved by HGGA1, but does correctly
recover the one-electron limit. (HGGA functionals achieving both properties
are currently under investigation in our group.) As Table~\ref{table2} 
shows, HGGA1$\cdot$MSIC combines better with exact exchange than any of the
other functionals, which is the behaviour expected from a HGGA. However, 
when combined with semilocal exchange it is still inferior to functionals 
exploiting error cancellation.

\section{Conclusions}
\label{concl}

All of the above suggests that the present HGGA functionals (with or without
the MSIC factor) deliver competitive correlation and atomization energies,
matching or outperforming those from sophisticated state-of-the-art functionals.
We stress, however, that in spite of this encouraging conclusion we consider
the present HGGA functionals merely as illustrations of the use of our 
representations (\ref{rep1}) and (\ref{rep2}) in the construction and analysis
of functionals, and not as the final word in this regard. 

Future development of
other functionals based on the same representations (including novel hybrids) 
seems promising. The way our HGGAs are constructed from the global Lieb-Oxford 
bound represents a radical departure from traditional modes of functional 
construction in quantum chemistry and solid-state physics \cite{dftbook,parryang,pw92,pbe,pw91,lyp,revpbe,hyb0,hyb2,hyb3,pkzb,tpss,PSTS,MCY,beckehgga}, 
which is only beginning to be explored.

This work was supported by FAPESP and CNPq.

\end{document}